\documentclass[aps]{revtex4}
\usepackage{amssymb}
\usepackage{graphicx}
\usepackage{bm}
\begin{document}
\def\vepsilon{{\hbox{\boldmath $\epsilon$ }}}

\title{On the derivation of an effective Higgs field}
\author{Hai-Jun Wang}
\address{Center for Theoretical Physics and School of Physics, Jilin
University, Changchun 130023, China}

\begin{abstract}
In one respect, the massive vector-boson shows its difference from
a massless vector-boson by one more physical polarization, known
as longitudinal polarization. In another respect, the quantized
boson acquires its mass by Higgs mechanism. In this paper we study
the effect of the longitudinal polarization in $U(1)$ case by
substituting it into the primary Yang-Mills Lagrangian $-\frac
14F_{\mu \nu }F^{\mu \nu }$. Under a hypothesis of strong
transversal condition for free vector boson, it is found that in
the Lagrangian the scalar field for the Higgs mechanism can
automatically arise after we separate a part equivalent to the
contribution of a massless boson. In addition, a criterion is
obtained to infer whether the boson is massive or not: if
$\mathbf{E}^2-\mathbf{B}^2\neq 0$, where $\mathbf{E}$ and
$\mathbf{B}$ are field strengths, then it is massive. The analysis
also pertains to $SU(2)$ case. The method in this paper is
performed before any quantizations.
\end{abstract}

\maketitle

The newly designed tera-scale Large Hadron Collider (LHC) of CERN is
expected to begin its switch within weeks ~\cite{1}~\cite{2}. One of its
missions is to search for Higgs particle. The "unexpected" results, which
push the possibility of the existence of Higgs particle to higher energy
scale, may appear at LHC. Then the conventional Higgs mechanism ~\cite{3}%
--which was thought to be accompanied by one light Higgs particle--has to be
modified. However, the light Higgs particle ($m_h\lesssim 144GeV$) meets the
stringent constraints by electroweak gauge couplings ~\cite{4}, and the most
recent experiments involving these constraints also point to a low value of
Higgs mass ~\cite{5}.

On the theoretical side, the Higgs mechanism is made a cornerstone related
to the vacuum in constructing a renormalizable theory with massive vector
bosons. It provides delicate cancellations of $\xi $-dependent
(gauge-fixing-condition dependent) terms in calculating $S$-matrix elements
of all orders. In spite of the success of Higgs mechanism, Veltman ~\cite{6}
insists that the scalar field required by a renormalizable theory should
have other origins if the light Higgs particle were not found. To seek
another scheme in place of Higgs mechanism, however, is not feasible at
present stage. Here we find a way out to validate Higgs mechanism without
the need of material Higgs field, in case that the light Higgs particle is
missing in experiment. The goal of this paper is to separate a term $%
\partial _\mu \varphi \partial ^\mu \varphi $ for vacuum states from the
free-field Lagrangian density $-\frac 14F_{\mu \nu }F^{\mu \nu }$ by
substituting the physical polarization vectors of massive boson.

The essence of Higgs mechanism includes two parts: one is the hypothesis
that there exists a scalar field responsible for the vacuum state (this
paper we name it \textbf{vacuum field}), which couples with vector bosons
via covariant derivative, $\partial _\mu \rightarrow \partial _\mu +ieA_\mu $%
; the other is, e.g. in $U(1)$ case, the specification of self-interaction
of the scalar field, which makes the global symmetry spontaneously broken
and meanwhile the scalar field gets a vacuum expectation value $\phi _0$.
Subtracted by the expectation value, the real part of the scalar field is
defined as the Higgs field. The complete form of Lagrangian involving Higgs
mechanism reads
\begin{equation}
\mathcal{L}=-\frac 14(F_{\mu \nu })_{massless}^2+|D_\mu \phi |^2-V(\phi ),
\end{equation}
in which $F_{\mu \nu }=\partial _\mu A_\nu -\partial _\nu A_\mu $
and $D_\mu =\partial _\mu +ieA_\mu $. For simplicity the subscript
"massless" means only two physical (spatially transversal)
polarizations of massless bosons are present in $A_\mu$. The
conventional decomposition of $\phi (x)$ is always used,
\begin{equation}
\phi (x)=\phi _0+\frac 1{\surd 2}(\phi _1(x)+i\phi _2(x)),
\end{equation}
where $\phi _1$ is Higgs field and $\phi _2$ is Goldstone field. In what
follows we mainly focus on the derivation of the term $|D_\mu \phi |^2$ in
Eq. (1) by substituting the massive polarization vectors into the primary
Yang-Mills Lagrangian $-\frac 14F_{\mu \nu }F^{\mu \nu }$.

In $U(1)$ case and for a massive neutral spin-1 vector boson which freely
moves along the third direction (later labelled as $z$ direction), there
exists a particular frame of reference where the three polarizations have
the form ~\cite{7}
\begin{eqnarray}
\mathbf{\epsilon }_1(\mathbf{k}) &=&(0,1,0,0)  \nonumber \\
\mathbf{\epsilon }_2(\mathbf{k}) &=&(0,0,1,0)  \nonumber \\
\mathbf{\epsilon }_3(\mathbf{k})
&=&(|\mathbf{k}|,0,0,E_{\mathbf{k}})/m,
\end{eqnarray}
here $k=(E_{\mathbf{k}},0,0,|\mathbf{k}|)$ and $k^2=k_\mu k^\mu =m^2$. The
polarization states meet the requirements of $k_\mu \mathbf{\epsilon }%
_{(}\lambda )^\mu (\mathbf{k})=0$ and $\mathbf{\epsilon }_{\lambda \mu }(%
\mathbf{k})\,\mathbf{\epsilon }_{\lambda ^{\prime }}^\mu (\mathbf{k}%
)=g_{\lambda \lambda ^{\prime }}$, here the summation convention (repeated
Greek indices are summed) is understood and the nonzero components of metric
tensor $g_{\lambda \lambda ^{\prime }}$ are $%
g_{00}=-g_{11}=-g_{22}=-g_{33}=+1$. Here the temporal polarization is absent
since for massive bosons only these three are ''physical'' ~\cite{7}.

The polarization vectors $\mathbf{\epsilon }_\lambda (\mathbf{k})$ and the
four-vector potential $A(t,\mathbf{x})$ ($\mathbf{x}=(x,y,z)$) are linked by
the Fourier expansion. We simplify the expansion by the above chosen frame
of reference, which is arbitrary but remains fixed temporarily. In this
frame, the plane wave has momentum $\mathbf{k}$,
\begin{eqnarray}
A(t,\mathbf{x}) &=&\int \frac{d\,\mathbf{k}}{\sqrt{2\omega _{\mathbf{k}%
}(2\pi )^3}}\sum_{\lambda =1}^3\mathbf{\epsilon }_\lambda (\mathbf{k}%
)(a_\lambda (\mathbf{k})e^{-ik\cdot x}+a_\lambda ^{*}(\mathbf{k})e^{ik\cdot
x})  \nonumber \\
&=&A_1(t,\mathbf{x},\mathbf{k})\mathbf{\epsilon }_1(\mathbf{k})+A_2(t,%
\mathbf{x},\mathbf{k})\mathbf{\epsilon }_2(\mathbf{k})+A_3(t,\mathbf{x},%
\mathbf{k})\mathbf{\epsilon }_3(\mathbf{k}),
\end{eqnarray}
where $\omega _{\mathbf{k}}=(m^2+\mathbf{k}^2)^{1/2}$. We will substitute
the Eq. (4) into the Lagrangian $-\frac 14F_{\mu \nu }F^{\mu \nu }$ to
compare the result with the case when two transversal polarization vectors $%
\mathbf{\epsilon }_1(\mathbf{k})$ and $\mathbf{\epsilon }_2(\mathbf{k})$
[These two are physical polarizations for photon] are present only, i.e. to
find the effect of $\mathbf{\epsilon }_3(\mathbf{k} )$ in $-\frac 14F_{\mu
\nu }F^{\mu \nu }$.

Since the Lagrangian $-\frac 14F_{\mu \nu }F^{\mu \nu }$ is a Lorentz
invariant quantity, it doesn't matter if we perform a Lorentz transformation
on four-vector potential $A_\mu (t,\mathbf{x})$. In view of the last line of
Eq. (4), the transformation will appear only to $\mathbf{\epsilon }_1(%
\mathbf{k})$, $\mathbf{\epsilon }_2(\mathbf{k})$ and $\mathbf{\epsilon }_3(%
\mathbf{k})$, and meanwhile the values of coefficients $A_\lambda (t,\mathbf{%
x},\mathbf{k}) $ will vary with $\mathbf{k}$, explicitly as $\frac{%
(a_\lambda (\mathbf{k})e^{-ik\cdot x}+a_\lambda ^{*}(\mathbf{k})e^{ik\cdot
x})}{\sqrt{2\omega _{\mathbf{k}}}}$. Let's suppose the system is boosted
along the $z$ direction with the velocity $\upsilon _z=\frac{|\mathbf{k}|}{%
E_{\mathbf{k}}}$(in natural unit, light velocity $c=1$) ~\cite{8}. This is
possible because the vector $\mathbf{\epsilon }_3(\mathbf{k})$ is space-like
$(\frac{E_{\mathbf{k}}}m)^2-(\frac{|\mathbf{k}|}m)^2=1>0$. Then in the new
reference frame
\begin{equation}
|\mathbf{k}^{\prime }|=\frac{|\mathbf{k}|-\upsilon _zE_{\mathbf{k}}}{\sqrt{%
1-\upsilon _z^2}},\;E_{\mathbf{k}}^{\prime }=\frac{E_{\mathbf{k}}-\upsilon
_z|\mathbf{k}|}{\sqrt{1-\upsilon _z^2}}\text{ .}
\end{equation}
Note that $|\mathbf{k}|=\upsilon _zE_{\mathbf{k}}$, one obtains
\begin{equation}
E_{\mathbf{k}}^{\prime }=\frac{E_{\mathbf{k}}-\upsilon _z^2E_{\mathbf{k}}}{%
\sqrt{1-\upsilon _z^2}}=\sqrt{E_{\mathbf{k}}^2-|\mathbf{k}|^2}=m\text{ ,}
\end{equation}
and thus

\begin{equation}
\mathbf{\epsilon }_L^{\prime }(\mathbf{k})=(0,0,0,1)\text{ .}
\end{equation}
Whereas under this transformation the other two polarization vectors $%
\mathbf{\epsilon }_1(\mathbf{k})$ and $\mathbf{\epsilon }_2(\mathbf{k})$
remain unchanged. In view of Eq. (7) and the fact that massless vector boson
owns only two physical polarizations, one concludes that \textbf{the
massless vector fields are not the limit of zero mass of massive vector
fields} (which is also mentioned in another manner in ~\cite{6}), since in
Eq. (7) the longitudinal polarization is mass independent. The potential $%
A_\mu $ in Eq. (4) now yields
\begin{equation}
A^{\prime }(t,\mathbf{x})=A_1^{\prime }(t,\mathbf{x},\mathbf{k})\mathbf{%
\epsilon }_1(\mathbf{k})+A_2^{\prime }(t,\mathbf{x},\mathbf{k})\mathbf{%
\epsilon }_2(\mathbf{k})+A_3^{\prime }(t,\mathbf{x},\mathbf{k})\mathbf{%
\epsilon }_L^{\prime }(\mathbf{k})\text{ .}
\end{equation}

Next let's examine how the third polarization (i.e. $A_3$ or $A_z$) in
Eq.(7) affects the Lagrangian $-\frac 14(F_{\mu \nu })^2$. The $U(1)$
field-strength tensor has the following form
\begin{eqnarray}
F^{\mu \nu } &=&\left(
\begin{array}{cccc}
0 & \frac \partial {\partial x}A_t-\frac \partial {\partial t}A_x & \frac
\partial {\partial y}A_t-\frac \partial {\partial t}A_y & \frac \partial
{\partial z}A_t-\frac \partial {\partial t}A_z \\
-\frac \partial {\partial x}A_t+\frac \partial {\partial t}A_x & 0 & \frac
\partial {\partial y}A_x-\frac \partial {\partial x}A_y & \frac \partial
{\partial z}A_x-\frac \partial {\partial x}A_z \\
-\frac \partial {\partial y}A_t+\frac \partial {\partial t}A_y & -\frac
\partial {\partial y}A_x+\frac \partial {\partial x}A_y & 0 & \frac \partial
{\partial z}A_y-\frac \partial {\partial y}A_z \\
-\frac \partial {\partial z}A_t+\frac \partial {\partial t}A_z & -\frac
\partial {\partial z}A_x+\frac \partial {\partial x}A_z & -\frac \partial
{\partial z}A_y+\frac \partial {\partial y}A_z & 0
\end{array}
\right)   \nonumber \\
&=&\left(
\begin{array}{cccc}
0 & E_x & E_y & E_z \\
-E_x & 0 & B_z & -B_y \\
-E_y & -B_z & 0 & B_x \\
-E_z & B_y & -B_x & 0
\end{array}
\right) ,
\end{eqnarray}
in which $\mu $ represents the lines and $\nu $ denotes the columns.
Substituting the transformed potential $A_\mu ^{\prime }$ in Eq.(8) [In what
follows we will still refer to $A_\mu $ as $A_\mu ^{\prime }$, with Eq.(7)
as the longitudinal polarization] into Eq.(9), it is found that the terms
including $A_z$ appear only in the last line and the last column of the
matrix. Considering the following strong transversal condition for free
vector boson
\begin{eqnarray}
\frac \partial {\partial z}A_x=\frac \partial {\partial z}A_y=\frac \partial
{\partial z}A_z=0,
\end{eqnarray}
(where we introduce the condition$\frac \partial {\partial
z}A_x=\frac
\partial {\partial z}A_y=0$ phenomenologically
and $\frac \partial {\partial z}A_z=0$ is in coincidence with the
Coulomb gauge fixing condition $\mathbf{k\cdot A=0}$. In
developing this paper, we have done our best not to be involved in
the using of gauge fixing conditions, which are closely related
with the quantization of fields) and substituting the component
form of field strengths $E_z=-\frac \partial {\partial t}A_z$,
$B_y=\frac
\partial {\partial x}A_z$, $B_x=-\frac \partial {\partial y}A_z$ into $%
F^{\mu \nu }$ and subsequently into $-\frac 14(F_{\mu \nu })_{massive}^2$,
we obtain the relevant terms including $A_z$ as follows
\begin{eqnarray}
\delta \mathcal{L}=-\frac 14(F_{\mu \nu })_{massive}^2(\text{relevant parts}%
)=\frac 12(E_z^2-B_y^2-B_x^2)\text{ .}
\end{eqnarray}
If we introduce a scalar field $\varphi =\frac 1{\sqrt{2}}A_z(t,\mathbf{x},%
\mathbf{k})$, then the terms in $\delta \mathcal{L}$ can be
straightforwardly written as
\begin{eqnarray}
\delta \mathcal{L}=(\partial _t\varphi )^2-(\partial _x\varphi )^2-(\partial
_y\varphi )^2=\partial _\mu \varphi \partial ^\mu \varphi \text{ .}
\end{eqnarray}
By this way one obtains the term $\partial _\mu \varphi \partial ^\mu
\varphi $ in the Lagrangian for Higgs mechanism. The separation of this term
requires that the component $A_z(t,\mathbf{x},\mathbf{k})$ should not be
transformed to zero by gauge transformation, therefore the separation
procedure here is by no means gauge-invariant.

So far the original Lagrangian $-\frac 14(F_{\mu \nu })_{massive}^2$ falls
into two parts:
\begin{eqnarray}
\mathcal{L}_{massive}=-\frac 14(F_{\mu \nu })_{massive}^2=\mathcal{L}%
_{massless}+\delta \mathcal{L},
\end{eqnarray}
where the first term of right hand side is the same as that for massless
photon, in which only two polarization vectors $\mathbf{\epsilon }_1(\mathbf{%
k})$ and $\mathbf{\epsilon }_2(\mathbf{k})$ are present. In general one puts
another two polarization vectors (longitudinal and scalar) additional to
them to form complete polarization states of massless bosons (then make it
mixed by Lorentz transformations). In scattering processes the physical
effects of redundant polarizations are cancelled ~ \cite{7}. Let's add a
self-interaction term $V(\varphi )$ [somehow produced by vacuum] to Eq.(13),
then the original Lagrangian for massive boson becomes
\begin{equation}
\mathcal{L}=-\frac 14(F_{\mu \nu })_{massless}^2+\partial _\mu \varphi
(x)\partial ^\mu \varphi (x)-V(\varphi )\text{ .}
\end{equation}
We can see that the Lagrangian already takes the form that for Higgs
mechanism in Eq. (1), with $\varphi$ identical with $\phi$, if only we make
replacement $\partial _\mu \rightarrow \partial _\mu +ieA_\mu $ to gain the
coupling between $\varphi (x)$ and $A_\mu $ as usual. To use the Higgs
mechanism, the next step is to make the \textbf{vacuum field} $\varphi $
complex to include a part responsible for the Goldstone field, as shown in
Eq. (2). Finally we should turn to the application of Higgs mechanism and
the corresponding quantization etc..

Noting the physical fact that the value $\mathcal{L}_{massless}=
\mathbf{E}^2-\mathbf{B}^2=0$, we recognize that in Eq. (13)
$\mathcal{L} _{massive}$ must be nontrivial, i.e.
$\mathcal{L}_{massive}=\mathbf{\tilde E}^2-\mathbf{ \tilde
B}^2\neq 0$ , otherwise the existence of $\mathbf{\epsilon }_3(
\mathbf{k})$ becomes nonsense [as usual, here
$E^i\mathbf{=}F^{i0}$ and $ B^i=-\varepsilon ^{ijk}F^{jk}$ as
default form for the massless cases, with $ \mathbf{\tilde E}$ and
$\mathbf{\tilde B}$ the same form but for massive cases].
Moreover, one should note that by reversing the above derivation,
the field $\varphi (x)$ turns out to be a component of $A_\mu $,
gained by the aid of Lorentz transformation, and dependent on the
breaking of gauge transformation.

The above separation steps are applicable to $SU(2)$ case too. The
definition of $F_{\mu \nu }$ for non-Abelian gauge fields is
\begin{equation}
F_{\mu \nu }^a=\partial _\mu A_\nu ^a-\partial _\nu A_\mu ^a+g\varepsilon
^{abc}A_\mu ^bA_\nu ^c,
\end{equation}
with $g$ as coupling constant and indices $a$ corresponding to the
generators of gauge group. By performing the gauge transformation
\begin{eqnarray}
A_\mu ^a &\rightarrow &A_\mu ^a+\frac 1g(\partial _\mu \alpha
^a)+f^{abc}A_\mu ^b\alpha ^c,  \nonumber \\
&&(\text{$f^{abc}$ are structure constants for gauge group})
\end{eqnarray}
the last term in Eq. (18) can be transformed away by gauge, and
thus the massive Lagrangian $-\frac 14F_{a\mu \nu }F^{a\mu \nu }$
reads ~\cite{9}
\begin{eqnarray}
-\frac 14F_{a\mu \nu }F^{a\mu \nu } &\rightarrow &-\frac 14\widetilde{F}%
_{a\mu \nu }\widetilde{F}^{a\mu \nu }  \nonumber \\
\text{with }\widetilde{F}_{\mu \nu }^a &=&\partial _\mu A_\nu ^a-\partial
_\nu A_\mu ^a\text{ ,}
\end{eqnarray}
which is analogous to the $U(1)$ case. Now we can perform the
similar separation Eq. (7) $\sim$ Eq. (13) in $U(1)$ case on Eq.
(17), the difference is that now the field strengths
$E^i\mathbf{=}\widetilde{F}^{i0}$ and $ B^i=-\varepsilon
^{ijk}\widetilde{F}^{jk}$ are all matrices. For $SU(2) $ case
three scalar fields $\varphi ^a$ $[a=1,2,3]$ are acquired. But to
fit the conventional application, we assume they are equal. Again
we note that this procedure is not gauge invariant due to the
derivation from Eq. (16) to Eq. (17).

In the $SU(2)$ case we note that the field $\varphi $ is also a component of
the field $A_\mu $,
\begin{equation}
\varphi =\frac 1{\sqrt{2}}A_z(t,\mathbf{x},\mathbf{k})\text{ ,}
\end{equation}
as in $U(1)$ case. However, since the field $A_\mu $ is a matrix, $A_\mu
^a\tau ^a$ ($\tau ^a=\frac{\sigma ^a}2$, with $\sigma ^a$ the Pauli
matrices), when we use the covariant differential $D_\mu =\partial _\mu
+ieA_\mu $ instead of $\partial _\mu $ to produce the coupling between $%
\varphi $ and $A_\mu $, at the first sight it seems problematic to make $%
\varphi $ act as a $SU(2) $ scalar $\left(
\begin{array}{c}
\eta _1+i\,\eta _2 \\
\upsilon +\sigma (x)+i\,\eta _3
\end{array}
\right) $ in the sense of the $SU(2)\otimes U(1)$ standard model. Whereas we
find the effect of $\varphi $ is identical with that of $SU(2)$ scalar $%
\left(
\begin{array}{c}
0 \\
\upsilon
\end{array}
\right) $ if one reduces the summation $A_z^a\tau ^a$ (over the indices $a$)
to a particular term, e. g. $A_z^3\tau ^3$. Let's Recall the mass-term
production in Higgs mechanism for $SU(2)$ case, where we replace the
differential $\partial _\mu $ in $\partial _\mu \varphi \partial ^\mu
\varphi $ with covariant one $D_\mu \varphi =(\partial _\mu -igA_\mu ^a\tau
^a)\varphi $, and make $\varphi =\left(
\begin{array}{c}
0 \\
\upsilon
\end{array}
\right) $, then the relevant part of the Lagrangian yields ~\cite{10}

\begin{eqnarray}
\triangle \mathcal{L} &=&\frac 12\varphi (g\,A_\mu ^a\tau ^a)\,(g\,A^{\mu
b}\tau ^b)\varphi  \nonumber \\
&=&\frac 12(0,\upsilon )(g\,A_\mu ^a\tau ^a)\,(g\,A^{\mu b}\tau ^b)\left(
\begin{array}{c}
0 \\
\upsilon
\end{array}
\right)  \nonumber \\
&=&\frac 12\frac{\upsilon ^2g^2}4A_\mu A^\mu \text{ ,}
\end{eqnarray}
by which $A_\mu $ gains a mass $\frac{g\upsilon }2$. Now the similar
expression exists even if the field $\varphi $ takes a matrix form $\frac 1{%
\sqrt{2}}A_z^3\tau ^3$, as follows
\begin{eqnarray}
\triangle \mathcal{L} &&=\frac 12\varphi (g\,A_\mu ^a\tau ^a)\,(g\,A^{\mu
b}\tau ^b)\varphi  \nonumber \\
&=&\frac 12\frac 1{\sqrt{2}}\frac 12\left(
\begin{array}{cc}
1 & 0 \\
0 & -1
\end{array}
\right) A_z^3(g\,A_\mu ^a\tau ^a)\,(g\,A^{\mu b}\tau ^b)\frac 1{\sqrt{2}%
}\frac 12\left(
\begin{array}{cc}
1 & 0 \\
0 & -1
\end{array}
\right) A_z^3  \nonumber \\
&=&\frac 1{64}(A_z^3)^2g^2\left(
\begin{array}{cc}
1 & 0 \\
0 & -1
\end{array}
\right) [(A_\mu ^1)^2+(A_\mu ^2)^2+(A_\mu ^3)^2]\left(
\begin{array}{cc}
1 & 0 \\
0 & -1
\end{array}
\right)  \nonumber \\
&=&\frac 1{64}(A_z^3)^2g^2A_\mu A^\mu ,
\end{eqnarray}
by this way the field $A_\mu $ gains a mass proportional to $(A_z^3)^2g^2$.
So far making the matrix form in place of the original $SU(2)$ scalar field
hasn't induced any problem in producing boson's mass. However, this scheme
seems not appropriate to the $SU(3)$ case since $(g\,A_\mu ^a\lambda
^a)\,(g\,A^{\mu b}\lambda ^b)$ ($\lambda ^a$ is Gell-Mann matrices) cannot
be written as $(A_\mu ^1)^2+(A_\mu ^2)^2+\cdot \cdot \cdot +(A_\mu ^8)^2$
without cross terms.

In summary, in this note we find the \textbf{vacuum field}, and thus the
Higgs field can be separated from the Lagrangian for massive boson field,
whence we may feel easy even if at LHC the Higgs particle were not claimed
after the near-future collection of data. From the separation procedure, one
notes that part of the spontaneous breaking of symmetries in conventional
Higgs mechanism has been transferred to the breaking of $U(1)$ or $SU(2)$
gauge symmetries before using the Higgs mechanism. So the separation we
performed is by no means a gauge invariant procedure. However it is
feasible---we have found where the energy of the vacuum states possibly
hides in the massive Lagrangian $-\frac 14(F_{\mu \nu })_{massive}^2$. The
result of the decomposition of massive-boson's Lagrangian provides us a
criterion to judge whether a boson owns mass or not: If $\mathbf{\tilde E}^2-%
\mathbf{\tilde B}^2\neq 0$, then it does. We choose the field
strengths $\mathbf{\tilde E}$ and $ \mathbf{\tilde B}$ to express
this criterion for they are observables even in classical cases.
Furthermore, we conclude from Eq. (20) that the boson's mass is
born from the self-interaction. Throughout this paper we have
worked in physical polarizations, regardless of the completeness
of them. In addition, the steps of separation happens before any
quantizations are performed, so we don't get involved in any
particular gauge fixing conditions, as well as the corresponding
propagator or other Green functions. Consequently, the field
strengths $\mathbf{\tilde E}$ and $ \mathbf{\tilde B}$ are not
operators yet.

The original motivation of this paper was inspired by the Goldstone
equivalence theorem~\cite{10} and one of the results of the manuscript~\cite
{11}. This work is supported by National Natural Science Foundation of China
under Grant Nos. 10675054 and 10775059.


\begin{thebibliography}{99}
\bibitem{8}  Breit system: When a four-vector is space-like, we can always
choose a reference frame in which one of its component is transformed away.
\end{thebibliography}
\end{document}